\documentstyle[array,aps,psfig]{revtex}
\newcommand{\Bibitem}[1]{\bibitem{#1}}        
\newcommand{\be}{\begin{equation}}
\newcommand{\ee}{\end{equation}}
\newcommand{\ba}{\begin{eqnarray}}
\newcommand{\ea}{\end{eqnarray}}
\newcommand{\nn}{\nonumber\\}
\newcommand{\Ref}[1]{(\ref{#1})}
\newcommand{\av}[1]{\langle #1\rangle}  
\newcommand{\half}{\textstyle{\frac{1}{2}}}

\newcommand{\fourth}{\textstyle{\frac{1}{4}}}
\newcommand{\al}{{\alpha}}

\newcommand{\tbkmin}{\tilde{{\bf k}}_-}
\newcommand{\tbkplus}{\tilde{{\bf k}}_+}

\newcommand{\tf}{\tilde{f}}
\newcommand{\tphi}{\tilde{\phi}}

\newcommand{\bk}{{\bf k}}

\newcommand{\bn}{{\bf n}}

\newcommand{\bv}{{\bf v}}
\newcommand{\bc}{{\bf c}}

\newcommand{\xovery}[2]{\left( \begin{array}{c} #1 \\ #2\end{array} \right)}

\begin{document}

\title{Scaling Solutions of Inelastic Boltzmann Equations \\ with
Over-populated  High Energy Tails\footnote{This article is
dedicated to our dear friend Robert J. Dorfman in honor of his
65-th birthday.}}
\author{M. H. Ernst\\
Instituut voor Theoretische Fysica\\
Universiteit Utrecht, Postbus 80.195\\ 3508 TD Utrecht, The
Netherlands\\[1mm]
and R. Brito\\
Departamento de F\'{\i}sica Aplicada I\\ Universidad Complutense\\
28040 Madrid, Spain}

\date{\today}
\maketitle

\vspace{2cm}

\noindent
Pacs: 05.20.Dd (Kinetic theory), 05.20.-y (Classical Statistical
Mechanics), 45.70.M (Granular Systems) \\


\begin{abstract}
This paper deals with solutions of the nonlinear Boltzmann
equation for spatially uniform freely cooling inelastic Maxwell
models for large times and for large velocities, and the
nonuniform convergence to these limits. We demonstrate how the
velocity distribution approaches in the scaling limit to a
similarity solution with a power law tail for general classes of
initial conditions and derive a transcendental equation from which
the exponents in the tails can be calculated. Moreover on the
basis of the available analytic and numerical results for
inelastic hard spheres and inelastic Maxwell models we formulate
a conjecture on the approach of the velocity distribution
function to a scaling form.
\end{abstract}

\renewcommand{\theequation}{I.\arabic{equation}}
\setcounter{section}{0} \setcounter{equation}{0}
\section{Introduction}

In recent times overpopulation  of high energy times in velocity
distributions of freely cooling or in driven granular fluids has
become a focus of attention in laboratory experiments, kinetic
theory and computer simulations \cite{physics-today}. In kinetic
theory, granular fluids out of equilibrium were originally
modeled by inelastic hard spheres
(IHS)\cite{campbell,goldhirsch}, which is the proto-typical model
for dissipative short ranged hard core interactions. In general,
similarity solutions are of interest, because they play an
important role as asymptotic or limiting solutions of the
Boltzmann equation at large times or at large velocities, and
they frequently show overpopulated high energy tails when
compared to the omni-present Gaussians.

Overpopulated tails were first found theoretically by studying
scaling or similarity solutions of the nonlinear Enskog-Boltzmann
equation for the IHS fluid, both in  freely evolving IHS systems
without energy input \cite{twan-gran-mat}, as well as in driven or
fluidized systems \cite{twan-gran-mat,piasecki,biben+trizac}, and
confirmed afterwards by Monte Carlo simulations of the Boltzmann
equation \cite{brey-DSMC,santos}, and by laboratory
experiments\cite{physics-today}. Overpopulated tails in free IHS
fluids \cite{brito+huthman} and driven ones\cite{ignacio-I+II} 
have also been studied by molecular dynamics simulations of
inelastic hard spheres. The observed overpopulations in IHS
systems are mainly stretched exponentials $\exp[-Ac^b]$ with $b=1$
\cite{esipov,twan-gran-mat,brey-DSMC,santos} in systems without
energy input, and $b=3/2$\cite{twan-gran-mat,santos} in driven
systems, but for some forms of driving\cite{santos,piasecki}
$b=2$ has been observed. For hard sphere systems there is in
general good agreement between the analytic predictions and
numerical or Monte Carlo solutions of the nonlinear Boltzmann
equation.

About two years ago Boltzmann equations for inelastic Maxwell
models have been introduced for the free case without energy
input in Ref. \cite{Ben-Naim,bobylev}, and for the driven case in
Ref. \cite{cercignani,b,c}. In these studies similarity solutions
have received a great deal of attention, because general classes
of such solutions turned out to be non-positive, and hence
unphysical.

The interest in overpopulated tails in {\em elastic} Maxwell
models has already a long history, and originated from the
discovery of an exact positive similarity solution of the
nonlinear Boltzmann equation for Maxwell molecules, the so-call
BKW mode \cite{krupp,bobylev-bgk,krook+wu,phys-rep-ME}, named
after Bobylev, Krook and Wu. The most recent interest in
similarity solutions of the Boltzmann equation for inelastic
Maxwell models (IMM) was also stimulated by the discovery of an
exact similarity solution \cite{Baldassarri-1D} for a freely
cooling one-dimensional IMM of the form $\tf(c)=(2/\pi)
1/[1+c^2]^2$. It is positive, has finite mean energy $\av{c^2}$,
and an algebraic high energy tail $1/c^4$. The same authors also
obtained \cite{Baldassarri-1D,baldassarri-private} Monte Carlo
solutions $f(v,t)$ of the nonlinear Boltzmann equation for freely
evolving IMM systems with general initial distributions both in
one and two dimensions, and observed that their numerical results
for $f(v,t)$ at large time could be collapsed on a scaling form
$f(v,t) \sim (v_o(t))^{-d} \tf( v/v_o(t))$. Here $v^2_o(t) \sim
\av{v^2} \sim \exp[- \gamma t]$, where the decay rate $\gamma
\sim (1-\alpha^2)$, represents the typical decay of the average
kinetic energy. Moreover, these scaling solutions showed heavily
overpopulated power law tails with a cross-over time, $\tau_{{\rm
tail}}(v)$, that increases with the energy, and $\alpha$ is the
coefficient of normal restitution.

Soon after that Krapivsky and Ben-Naim \cite{BN+PK-6-11} and the
present authors \cite{ME+RB}, using a self-consistent method,
gave a theoretical explanation of these power law tails for
general dimensionality $d$ together with explicit predictions of
the tail exponent $a$, defined through $\tf(c) \sim 1/c^{2a+d}$.
These results were recently extended \cite{ME+RB-rapid,nienhuis} to
driven inelastic Maxwell models, where the high energy tail is of
exponential form $~ \exp[ -A|c|]$. Apparently, the type of
overpopulation depends sensitively on the microscopic model, on
the degree of inelasticity and  on the possible mode of energy
supply to the dissipative system.

    From the point of view of kinetic theory the intriguing question
is, what is the generic feature causing overpopulation of high
energy tails  in systems with {\em inelastic} particles, rather
than what is the specific shape of the tail. How does the
overpopulation depend on the underlying microscopic model, and on
the different forms of energy input
\cite{santos,piasecki,cercignani,b,c}? As discussed more
extensively in Ref. \cite{ME+RB-rapid} the differences in shape
are frequently related to non-uniformities in the limits of long
times, large velocities and vanishing inelasticity, which lead to
different results when taking the limits in different order or
when taking coupled limits such as the scaling limit (e.g. the
differences between bulk and tail behavior), or performing an
expansion in powers of the inelasticity, and then studying large
times (typically Gaussian tails are observed\cite{cercignani}) or
studying large times at fixed inelasticity and taking large time
limits afterwards (typically overpopulated tails are observed
\cite{piasecki,biben+trizac}) with a whole wealth of coupled
limits in between.

What are inelastic Maxwell models? The IMM's, introduced in
\cite{Ben-Naim,bobylev}, share with elastic Maxwell molecules the
property that the collision rate in the Boltzmann equation is
independent of the relative kinetic energy of the colliding pair.
However, these IMM's do not describe real particles, but only
pseudo-particles with 'collision rules' $(\bv_1,\bv_2) \to
(\bv^*_1,\bv^*_2)$ between pre- and post-collision velocities,
defined to be the same as for smooth inelastic hard spheres with
a restitution coefficient $\alpha$ with $ (0<\alpha<1)$.  There
are no such objects as 'inelastic Maxwell particles' that
interact according to a given force law and that can be studied
by molecular dynamics simulations.

These IMM's are of interest for granular fluids in spatially
homogeneous states, not because they can claim to be more
realistic than IHS's, but because of the mathematical
simplifications resulting from an energy-independent collision
rate. Nevertheless the IMM's keep  the qualitatively correct
structure and properties of the nonlinear macroscopic equations
\cite{G+Z} and obey Haff's law \cite{haff}, just like the even
simpler inelastic BGK or single relaxation time models
\cite{brey-dufty-santos} do. What harmonic oscillators are for
quantum mechanics, and dumb-bells for polymer physics, that is
what elastic and inelastic Maxwell models are for kinetic theory.

    From the point of view of nonequilibrium (steady) states, the
structure of velocity distributions in dissipative systems,
including the high energy tail, is a subject of continuing
research, as the universality of the Gibbs' state of thermal
equilibrium is lacking outside thermal equilibrium, and a
possible classification of generic structures would   be of  great
interest in many fields of non-equilibrium statistical mechanics.

After this explanation of the possible relevance of inelastic
Maxwell models for different fields of research, we concentrate on
the kinetic theory for these models, in particular on the
simplest case, the freely cooling one without energy input. We
return now to the recent  results of
Refs.\cite{Baldassarri-1D,Baldassarri,baldassarri-private}, and we
consider their observations, described above, as strong evidence
for the existence of interesting limiting behavior when coupled
limits are taken, and more explicitly, we interpret their findings
as follows: the transformed or {\em rescaled} distribution
function $\tf(c,t)$ defined through,
\be \label{def-f-ct}
 f(v,t) = (v_o(t))^{-d} \tf(v/v_o(t),t),
\ee
approaches a scaling or similarity form in the coupled limit as
$t\to \infty$ and $v\to 0$, with $v/v_o(t)=c$ kept constant, i.e.
\be \label{def-f-c}
 \lim_{t \to \infty} \tilde{f} (c,t) =
\lim_{t \to \infty} (v_o(t))^d f(v_o(t)c,t)= \tilde{f}(c).
\ee
The coupled limit considered in (\ref{def-f-c}) is  the same
scaling limit as considered in the solutions  of the nonlinear
Boltzmann equation for IHS in
Refs.\cite{twan-gran-mat,ME+RB-rapid}, although it has not been
pointed out so emphatically. On the other hand the scaling limit,
considered  in Refs. \cite{bobylev,cercignani,b,c}, is the
coupled limit as $t \to \infty$ and $\alpha \to 1$ with
$(1-\alpha^2)t= const$, while $v$ is kept constant.

The present paper leads to the formulation of a {\em conjecture}
by combining the recent results on scaling solutions and
overpopulated high energy tails in inelastic hard sphere fluids
and inelastic Maxwell models  with an older conjecture of Krook
and Wu \cite{krook+wu} on the role of a special self-similar
solution, the so-called BKW-mode
\cite{krupp,bobylev-bgk,krook+wu,phys-rep-ME}, named after
Bobylev, Krook and Wu. This conjecture reads: {\em  Solutions
$f(v,t)$ of the nonlinear Boltzmann equation for dissipative
systems -- i.e. the rescaled distribution $\tf(c,t)$ in {\rm
\Ref{def-f-ct}} rescaled with the instantaneous r.m.s. velocity
$v_o(t) \sim (\av{v^2})^{1/2}$ -- approach for general initial
conditions, in the scaling limit {\rm \Ref{def-f-c}}, to the scaling
solution $\tf(c)$ with an overpopulated high energy tail. In
taking the scaling limit the degree of inelasticity,
$(1-\alpha^2)$, must be kept constant, and cannot be interchanged
with the elastic limit $(\al \uparrow 1)$}.

This conjecture is a variation on the Krook-Wu conjecture for
elastic Maxwell molecules, formulated as: "An arbitrary initial
state tends first to relax to a state, characterized by the
BKW-mode. The subsequent relaxation is essentially represented
the BKW-mode with an appropriate phase". As it turned out, this
conjecture was not supported by numerical and analytic results
obtained for the physically most relevant initial distributions
with a finite second moment in the limit as $t \to +
\infty$\cite{phys-rep-ME}. However, Bobylev and Cercignani have
recently shown that the conjecture holds in systems of elastic
Maxwell molecules in the limit $t \to - \infty$ for the so-called
{\em eternal} solutions $f(v,t)$\cite{AB+CCpreprint,B+C-sim-appl},
which are characterized by a divergent second moment.

The paper is organized as follows. In section II the mathematical
model Boltzmann equation for IMM's is constructed starting from
the Enskog-Boltzmann equation for IHS's, and some basic
properties are  derived there, as well as in Appendix A. In
section III we show that the IMM Boltzmann equation admits a
similarity solution  with a power law tail $\sim 1/c^{2a+d}$,
where the non-integer tail exponent $a$ is the solution of a
transcendental equation, which is solved numerically. The moments
$\av{c^{2n}}$  of the scaling form $\tf{(c)}$ with $n<a$ are
calculated in section IV.B from a recursion relation. The moments
with $n>a$ are divergent. In section IV.B  we demonstrate that
the moment $\mu_n(t)$ of the rescaled distribution $\tf(c,t)$,
for the general class of initial conditions with all moments
$\av{v^{2n}} < \infty$, approach in the long time limit for $n<a$
to the unique set of moments $\mu_n$ of the scaling form, and we
analyze how all moments $\mu_n(t) \to \infty$ for $n>a$. Some
generalizations of these results are described in section IV.C.
In section V we present our conclusions, and interpret our
results as a demonstration of our conjecture for IMM's, i.e. as a
weak form of approach of an arbitrary rescaled distribution
$\tf(c,t)$ to a universal scaling form.

\renewcommand{\theequation}{II.\arabic{equation}}
\setcounter{section}{1} \setcounter{equation}{0}
\section{Kinetic Equations for Dissipative Systems}
\subsection{Inelastic Hard Spheres }

For the construction of inelastic Maxwell models, it is convenient
to start from the spatially homogeneous Boltzmann equation for
inelastic hard spheres.  We study the velocity distribution,
$f(\bv,\tau)$ in the so-called homogeneous cooling state (HCS).
Here $\tau$ is the "external" laboratory time, and the relation to
time $t$ used in section I will be given in due time. Moreover we
restrict ourselves to isotropic distributions $f(v,\tau)$ with
$v=|\bv|$ with isotropic initial conditions, $f(v,0)$. The most
basic and most frequently used model for dissipative systems with
short range hard core repulsion is the Enskog-Boltzmann equation
for inelastic hard spheres in $d-$dimensions \cite{campbell},
\be \label{IHS-BE}
{\partial}_\tau f_1 = I(f) \equiv 2\int_\bn \int d\bv_2 \theta
({\bv}_{12} \cdot \bn )|{\bv}_{12} \cdot \bn |
[\textstyle{\frac{1}{\al^2}} f_1^{**} f_2^{**} -f_1 f_2],
\ee
where $f_1^{**}$ is short for $f(\bv_1^{**}, \tau)$, and  we have
absorbed constant factors in the time scale. Velocities and time
have been dimensionalized in terms of the width and the mean free
time of the initial distribution, and $|{\bv}_{12} \cdot \bn |  $
is essentially the dimensionless collision rate. Moreover,
$\int_\bn (\cdots) = (1/\Omega_d) \int d\bn(\cdots)$ is an
angular average over a $d$-dimensional unit sphere, restricted to
the hemisphere, ${\bv}_{12} \cdot \bn >0$, through the unit step
function $\theta (x)$, and $\Omega_d = 2 \pi^{d/2}/ \Gamma(\half
d)$.

The velocities ${\bv}_i^{**}$ with $i,j =\{1,2\}$ denote the {\em
restituting} velocities, and $\bv_i^{*}$ the corresponding {\em
direct} postcollision  velocities. They are defined as,
\ba \label{dyn}
\bv_i^{**}&=& \bv_i -\half
(1+\textstyle{\frac{1}{\al}})\bv_{ij}\cdot \bn \bn
\nn \bv_i^{*}& =& \bv_i -\half (1+\al)\bv_{ij}\cdot \bn \bn.
\ea
Here $\al$ is the coefficient of restitution $(0 < \al <1)$, the
relative velocity is $\bv_{ij} =\bv_i-\bv_j$, and $\bn$ is a unit
vector along the line of centers of the interacting particles. In
one dimension the angular average $\int_\bn$, as well as  the
dyadic product $\bn \bn$ can be replaced by the number  $1$. One
of the factors $(1/\al)$ in Eq.\Ref{IHS-BE}
 originates from the Jacobian, $d\bv_1^{**}
d\bv_2^{**} = (1/\al)d\bv_1 d\bv_2 $, and the other one from the
collision rate of the restituting collisions, $|{\bv}^{**}_{12}
\cdot \bn | = (1/\al) |{\bv}_{12} \cdot \bn |$. Furthermore, in
the HCS symmetrization over $\bn$ and $-\bn$ allows us to replace
$2\theta (x)$ in \Ref{IHS-BE} by $1$.

In subsequent sections we will also need the rate equations for
the average $\av{\psi}_\tau   =\int d \bv \psi (\bv)f(\bv,\tau)$,
as follows from the Boltzmann equation,
\be \label{rate-eq}
d\av{\psi}_\tau  /d\tau= \textstyle \int d\bv \psi (\bv) I(f) =
\int_\bn \int d \bv_1 d\bv_2|{\bv}_{12} \cdot \bn | f_1 f_2 [ \psi
(\bv^*_1) - \psi (\bv_1)].
\ee
The Boltzmann collision operator conserves the number of particles
$(\psi(\bv)=1)$ and momentum $(\psi(\bv)=\bv)$, but not the
energy $(\psi (\bv) = v^2)$. Here normalizations are chosen such
that,
\ba \label{norm}
\av{1}_\tau &= \int d\bv f(\bv,\tau) &=1
\nn \av{\bv}_\tau &= \int d\bv \bv f(\bv,\tau) &=0
\nn \av{v^2}_\tau &= \int d\bv v^2 f(\bv,\tau)&= \half d v_o^2(\tau).
\ea
 As a consequence of the inelasticity an amount of energy, $\fourth
(1-\al^2)[{\bv}_{12} \cdot \bn]^2$, is lost in every inelastic
collision. Consequently the average kinetic energy or granular
temperature $\av{v^2}$ keeps decreasing at a rate proportional to
the inelasticity $(1-\al^2)$. So, the solution of the Boltzmann
equation does not reach thermal equilibrium, described by the
Maxwellian $\varphi_o(\bv)= \pi^{-d/2} \exp[-v^2]$, but is
approaching a Dirac delta function $\delta^{(d)}(\bv)$ for large
times. As the convergence of $f(v,\tau)$ to its limiting value as
$\tau \to \infty$ is in general non-uniform, the rescaled
 distribution may approach a different limit
(see \Ref{def-f-c}). Also, the detailed balance condition is
violated, and the Boltzmann equation does not obey an
$H-$theorem. The moment equations and the behavior of the scaling
solutions for freely evolving and driven IHS fluids have been
extensively discussed both in the bulk of the thermal
distribution, as well as in the high energy tails
\cite{brey-DSMC,twan-gran-mat}.

\subsection{Inelastic Maxwell Models}
One of the difficulties in solving the nonlinear Boltzmann
equation \Ref{IHS-BE} for hard spheres is that the collision rate
$|\bv_{12} \cdot \bn|$ is not a constant, but is proportional to
the relative velocity $v_{12}$ of the colliding pair, which is
typically of order $ v_{12} \sim v_o(\tau)$, as defined in
(\ref{norm}). Maxwell models on the other hand are defined to have
a collision rate independent of the relative energy of the
colliding particles.

In the recent literature two different types of mathematical
simplifications have been introduced which convert the IHS-
Boltzmann equation into one for an inelastic Maxwell model with
an energy independent collision rate.  In the most drastic
simplification, the IMM-A discussed in Refs.
\cite{Baldassarri,BN+PK-6-11,ME+RB}, one replaces the collision
rate $|\bv_{12} \cdot \bn|$ for the direct collisions, as well as
the one for the restituting collisions, $|\bv^{**}_{12} \cdot
\bn| =|\bv_{12} \cdot \bn| / \al$ by its typical mean value
$v_o(\tau)$. In a more refined approximation Bobylev et al.
\cite{bobylev} replace these collision rates by $ v_o(\tau)
|\hat{\bv}_{12} \cdot \bn|$. We call this model IMM-B. Both
approximations keep the qualitatively correct dependence of the
total energy $v_o^2(\tau)$ on the 'external' time $\tau$. In
fact, strictly speaking these models should be called
pseudo-Maxwell molecules, because there do not exist microscopic
particles with dissipative interparticle forces, for which the
mathematical model Boltzmann equations below can be derived.

By making the above mathematical simplifications we obtain from
(\ref{IHS-BE}) a collision term which is multiplied by a factor
$v_o(\tau)$. This  factor is then absorbed by introducing a new
time variable $t$. For model IMM-A the resulting time
transformation and Boltzmann equation in dimensionless variables
are then given by,
\ba \label{IMMA}
dt &=& v_o(\tau)d\tau \nn
\partial_t f_1 &=& I(f) = \int_\bn \int d\bv_2 \left[
\textstyle{\frac{1}{\al}} f_1^{**}f_2^{**}-f_1 f_2 \right] \nn
\hspace{1cm} &=& -f_1 + \int_\bn \int d\bv_2
\textstyle{\frac{1}{\al}} f_1^{**}f_2^{**}.
\ea
For the IMM-B we introduce a slightly different time variable
$t$, and obtain the Boltzmann equation in dimensionless variables,
\ba \label{IMMB}
dt &=& \beta_1 v_o(\tau) d\tau \nn
\partial_t f_1 &=& I(f) = \int^\prime_\bn \int d\bv_2
 |\hat{\bv}_{12} \cdot \bn| \left[
\textstyle{\frac{1}{\al}} f_1^{**}f_2^{**}-f_1 f_2 \right] \nn
\hspace{1cm} &=& -f_1 + \int^\prime_\bn \int d\bv_2
|\hat{\bv}_{12} \cdot \bn| \textstyle{\frac{1}{\al}}
f_1^{**}f_2^{**},
\ea
where $\int^\prime_\bn= (1/\beta_1)\int_\bn$ with $\beta_1$
defined in \Ref{B1}. The prefactors in the time transformations
are chosen such that the loss term takes the simple form $-f_1$.
This implies that the new time variable $t(\tau)$ counts the
average number of collisions suffered by a particle within the
"external" time $\tau$. Hence $t$ is the collision counter or
"internal" time of a particle.

One of the important properties of Maxwell models is that the
moment equations form a set of coupled equations, that can be
solved {\em sequentially}. For the one-dimensional Maxwell model
these equations were derived in Ref.\cite{Ben-Naim}, and for the
three-dimensional model IMM-B with uniform impact parameter this
was done in Ref.\cite{bobylev}. The general moment equations for
the present Maxwell models  will be derived after having obtained
the characteristic function in Sect.III. At this point we make an
exception for the second moment, which determines the typical
velocity $v_o(t)$ through the relation $\av{v^2}_t =\half d
v_o^2(t)$,  needed to study the rescaled distribution function.
It follows from the kinetic equations \Ref{IMMA} and \Ref{IMMB} as
\ba \label{2-moment}
&\partial_t \av{v^2}_t &= - \gamma \av{v^2}_t \nn
 &\gamma &= \left \{ \begin{array} {ll}
      \frac{1-\al^2}{2d} =\frac{2p(1-p)}{d} &\qquad \mbox{(IMM-A)}
      \\[1mm]
     \frac{1-\al^2}{d+1}=\frac{4p(1-p)}{d+1} &\qquad \mbox{(IMM-B)}
\end{array}
\right.,
\ea
where $p=\half (1+\al)$.  Consequently $v_o(t) =
v_o(0)\exp[-\half \gamma t]$. Moreover, by solving the
differential equations for $t$ in \Ref{IMMA} and \Ref{IMMB} we
obtain the relations between the internal time $t$ and the
external time $\tau$, i.e.
\be \label{t-tau}
\exp[\half \gamma t] = \left\{ \begin{array}{ll}
     1+\half\gamma v_o(0) \tau & \qquad \mbox{(IMM-A)}\\[1mm]
     1+\half \beta_1\gamma v_o(0)\tau  &\qquad \mbox{(IMM-B)}
\end{array} \right.,
\ee
as well as the decay of the energy in terms of internal time $t$
and external time $\tau$, i.e.
\be \label{haff}
 v^2_o(t)=
 \exp[-  \gamma t]v^2_o(0)=\left\{\begin{array} {ll}
 v^2_o(0)/[1 +\half \gamma v_o(0) \tau]^2  &\qquad \mbox{(IMM-A)}\\[1mm]
v^2_o(0)/[1 +\half \beta_1 \gamma v_o(0) \tau]^2 & \qquad
\mbox{(IMM-B)}.
\end{array} \right..
\ee
This shows that Haff's law \cite{haff}, given by the second
equality, is also valid for inelastic Maxwell models.

 To further elucidate the difference between the two classes of
models we change the integration variables $\bn$ --- which
specifies the point of incidence on a $d$-dimensional action
sphere of two colliding particles --- to the impact parameter, $b=
|\hat{\bv}_{12} \times \bn| = \sin \theta$, where $\theta =
\cos^{-1}(\hat{\bv}_{12} \cdot \bn)$ is the angle of incidence.
The relevant structure of the integrals for IMM-A and IMM-B is
respectively,
\ba \label{Pb-A+B}
\int_\bn  &\sim \displaystyle \int^{\pi/2}_0 d\theta (\sin
\theta)^{d-2}  \sim \int^1_0 db b^{d-2} /\sqrt{1-b^2}&
\nn \int^\prime_{\bn} |\hat{\bv}_{12} \cdot \bn| &\displaystyle
\sim \int^{\pi/2}_0
d\theta (\sin \theta)^{d-2} \cos \theta   \sim \int^1_0 db
b^{d-2}&.
\ea
Therefore model A has a {\em uniform} distribution ${\cal P}
(\theta)=1$ over angles of incidence, and a non-uniform
distribution $P(b) = 1/\sqrt{1-b^2}$ over impact parameters,
which is biased towards grazing collisions, where $b=1$. Model B
has a uniform distribution $P(b)=1$ over impact parameters, and a
non-uniform distribution, ${\cal P}(\theta)=\cos \theta$, biased
towards zero angle of incidence. In this context it is important
to note that the arguments on the validity of the Boltzmann
equation are based on the assumption of {\em molecular chaos},
i.e. absence of precollision correlations between the velocities
of a colliding pair. This implies that the distribution function
of impact parameters be uniform. Hence, model IMM-A does not obey
molecular chaos.

The question of interest in then: do IMM-A and IMM-B yield
qualitatively the same results for the scaling distribution? The
question is relevant because Molecular Dynamics simulations (MD)
of a system of $N$ IHS have shown that the dissipative dynamics
\Ref{dyn} drives an initially uniform distribution $P(b)=1$ in
the HCS towards a non-uniform distribution $P(b)$, biased towards
grazing collisions, which violates molecular chaos. So the IMM-A
model with a built-in initial bias may lead to spurious effects,
such as power law tails in $f(v,t)$, which are artifacts of a too
drastic simplification.

\subsection{Similarity Solutions}
The questions addressed in this subsection are: do the Boltzmann
equations for the Maxwell models, constructed in the previous
section, admit similarity solutions, and what are the properties
of such solutions? We define a {\em similarity} solution $\tf(c)$
through the relation,
\be \label{sim-sol}
f(\bv,t) = v_o^{-d}(t) \tilde{f} (\bv/v_o(t)).
\ee
The normalizations  imposed by \Ref{norm} on these solutions are,
\be
\label{norm-scaled}
  \int d\bc \tilde{f}(\bc)=1 \qquad
  \int d\bc \; c^2 \tilde{f}(\bc)= \half d .
\ee
By inserting \Ref{sim-sol} in \Ref{IHS-BE}, and using $v_o(t)\sim
\exp[-\half \gamma t]$ we obtain the following integral equation
for $\tf(c)$, i.e.
\be\label{sim-BE}  \half{\gamma}
\frac{{\bf \partial}}{{\bf \partial}\bc} \cdot \bc \tilde{f}(c)=
\widetilde{I}(\tilde{f}).
\ee
Here the operator $\tilde{I}(\tilde{f})$ has the same functional
form as $I(f)$ in \Ref{IMMA} or \Ref{IMMB} with $\{\bv_i, f\}$
replaced by $\{\bc_i, \tilde{f}\}$.

One of the goals of this paper is also to analyze in section IV
in what sense the rescaled distribution function, $\tf(c,t)$,
approaches its limiting form as $t \to \infty$. To do so, we also
need the kinetic equation for the rescaled $\tf(c,t)$, which
reads,
\be   \label{t-scaled-BE}
\partial_t \tf + \half{\gamma} \frac{{\bf \partial}}{{\bf
\partial}\bc} \cdot \bc  \tf =  \tilde{I}(\tf).
\ee
Some comments are in order here. Physical solutions
$\tilde{f}(\bc)$ of \Ref{sim-BE} must be non-negative. A velocity
distribution $f(\bv,t)$, evolving under the nonlinear Boltzmann
equation, preserves  positivity  for a positive initial
distribution $f(\bv,0)$ \cite{resibois-book,cercignani-book}.
However, for scaling solutions, being the solution of
\Ref{sim-BE}, positivity is not guaranteed
\cite{bobylev,phys-rep-ME}. If one would know a positive scaling
solution $\tf(c)$ -- as is the case in one dimension
\cite{Baldassarri-1D} -- and prepare the system in this initial
state, then the entropy $S$ or the $H-$function in state
\Ref{sim-sol}, also shows singular behavior, i.e.
\ba \label{H-attractor}
S(t) &=& - H(t) = - \int d\bc \tilde{f}(\bc) \ln{\tilde{f}(\bc)}
 -  \half d \gamma t +const \nn
 & \simeq & -\half d\,\gamma\, t + const \qquad ( t \; \;{\rm large}),
\ea
where $\gamma $ is positive  and $\int \tilde f\ln\tilde f$ is
some constant.  In these solutions the entropy keeps decreasing
at a constant rate $\half d\;\gamma$. This is typical for pattern
forming mechanisms in configuration space, where spatial order or
correlations are building up, as well as in dynamical systems and
chaos theory, where the rate of irreversible entropy production
is negative on an attractor \cite{Ruelle,Evans,Dorfman}. In fact the
forward  dissipative dynamics, $(\bv_1,\bv_2) \to (\bv_1^{*},
\bv_2^{*})$, defined in \Ref{dyn}, has a Jacobian $J=\al <1$, i.e.
$d\bv_1^{*} d\bv_2^{*} = \al d\bv_1 d\bv_2 $, corresponding to a
contracting flow in $\bv-$space. Moreover, there is no
fundamental objection against decreasing entropies in an open
subsystem, here the inelastic Maxwell particles, interacting with
a reservoir. The reservoir is here the sink, formed by the
dissipative collisions, causing the probability to contract onto
an attractor.

\renewcommand{\theequation}{III.\arabic{equation}}
\setcounter{section}{2} \setcounter{equation}{0}
\section{Power Law Tails}
\subsection{Fourier transformed Boltzmann equation}

The goal of this section is to show that the Boltzmann equation
for IMM's has a scaling solution with a power law tail. This is
done by introducing the Fourier transform of the distribution
function, $\varphi(\bk,t) =  \av{\exp[-i\bk \cdot {\bv}]}_t$,
which is the characteristic function or generating function of the
velocity moments. Because $f(v,t)$ is isotropic, $\varphi(k,t)$
is isotropic as well. It is also convenient to consider
$\phi(x,t)$, defined through the relation $\varphi (k,t) =
\phi(\fourth k^2,t)$.

We start with the simplest case, and apply Bobylev's Fourier
transform method \cite{bobylev,bobylev-bgk} to the Boltzmann
equation \Ref{IMMA} for model IMM-A with the result,
\ba \label{ft-t-eq}
\partial_t \varphi(\bk,t)&=&
\int_\bn [\varphi(\bk_+,t)\varphi(\bk_{-},t)- \varphi({\bf
0},t)\varphi(\bk,t)] \nn
\partial_t \phi(x,t)&=&
\int_\bn [\phi(xe_+(\bn),t)\phi(xe_{-}(\bn),t)- \phi(
0,t)\phi(x,t)],
\ea
where $\varphi({\bf 0},t)=1=\phi(0,t)$. Here we have used
\Ref{rate-eq} with $\psi(\bv_1)=\exp[-i\bk\cdot\bv^*_1]$  and
expressed the exponent as $\bk\cdot \bv^*_1 = \bk_- \cdot \bv_1 +
\bk_+ \cdot \bv_2$ (see \Ref{dyn}), where
\ba \label{k-var}
\bk_{+} &\equiv k \tbkplus = p \bk \cdot \bn \bn  \qquad & \tilde
k_+^2 = p^2  (\hat{\bk}\cdot \bn)^2 = e_+(\bn)
\nn \bk_{-}& \equiv k \tbkmin   =\bk - \bk_{+} \qquad & \tilde k_-^2 =
 [ 1- z (\hat{\bk} \cdot \bn)^2] = e_-(\bn),
\ea
with $p=\half (1+\al)$ and $z=2p-p^2$. In one dimension this
equation simplifies to
\be \label{ft-1D-eq}
\partial_t \varphi(k,t)= \varphi(p k,t)\varphi((1-p)k,t)-\varphi(k,t),
\ee
where $k_+ =pk$ and $k_-=(1-p)k$. Equation \Ref{ft-t-eq} has the
interesting property that for a given solution $\varphi(\bk,t) $
one has a whole class of solutions $\overline{\varphi}(\bk,t) =
\exp[ i\bk \cdot {\bf w}]\varphi(\bk,t)$ where ${\bf w}$ is an
arbitrary velocity vector \cite{bobylev-bgk,bobylev}. This
property reflects the Galilean invariance of the Boltzmann
equation.

Because $f(v,t)$ is isotropic, only its even moments are
non-vanishing, and the moment expansion  of the characteristic
function  then takes the form,
\be \label{mom-t-exp}
\varphi(\bk,t)= \sum_{m=even} \frac{(-ik)^m}{m!} \langle
(\hat\bk\cdot\bv)^m\rangle_t =\sum_n \frac{(-x)^n}{n!}
m_n(t)=\phi(x,t),
\ee
where $x=k^2/4$. The angular time independent  average $\langle
(\hat\bk\cdot\hat\bv)^{2n}\rangle =\beta_{2n} $ is calculated in
\Ref{B1} and the moment $m_n(t)$ is defined as
\ba \label{mom-t-n}
 &m_n(t)=  4^n n! \beta_{2n}\av{v^{2n}}_t/(2n)!
 = \langle v^{2n}\rangle_t/(d/2)_n &
\nn & \beta_{2n}=\langle
(\hat\bk\cdot\hat\bv)^{2n}\rangle =(1/2)_n/(d/2)_n &.
\ea
The Pochhammer symbol $(a)_n$ is defined in \Ref{B3} and we have
used the duplication formula for the Gamma function
$(2n)!=\Gamma(2n+1)$. Furthermore we note that the moments of a
Gaussian $\varphi_o(v)=\pi^{-d/2} \exp(-v^2)$ are
$\av{v^{2n}}_o=(d/2)_n$.

Scaling solutions in Fourier representation have the form
$\phi(x,t)=\Phi(e_o(t)x)$, where $\Phi(\fourth k^2)$ is the
Fourier transform of $\tf(c)$ and $e_o(t)=v_o^2(t) = e^{-\gamma
t} e_o(0)$. Substitution in Eq.(\ref{ft-t-eq}) yields the integral
equation for the scaling form,
\be \label{ft-eq}
-\gamma x \Phi'(x)+\Phi(x) = \int_\bn \Phi(xe_+) \Phi(xe_-).
\ee
Its moments follow from the expansion,
\ba \label{mom-exp}
& \Phi (x) =\sum_n (-x)^n \mu_n/n! =1-x+x^2\mu_2/2!+\cdots & \nn
& \mu_n = \int d\bc c^{2n} \tf(c) /(d/2)_n \equiv
\av{c^{2n}}/(d/2)_n,
\ea
where (\ref{norm-scaled}) imposes $\mu_1=1$.

\subsection{Small-$k$ singularity of characteristic function}
In case all coefficients in the Taylor expansion \Ref{mom-exp}
exist, then $\Phi(x)$ is regular at the origin, and the
corresponding scaling form $\tf(c)$ falls off exponentially fast
at large $c$, and all its moments $\av{c^{2n}}$ are finite.
Suppose now that the small-$k$ or small-$x$ behavior of $\Phi(x)$
contains a singular term $ x^a$, where $a$ does not take integer
values (note that  even powers $k^{2a} =k^{2n} $ represent
contributions that are regular at small $k$), then its inverse
Fourier transform scales as $1/c^{2a+d}$ at large $c$. For this
distribution the moments with $n \geq a$ are divergent, and so is
the $n$-th derivative of the generating function $\Phi(x)$   at
$x=0$. The requirement that the total energy be finite imposes
the lower bound  $a>1$ on the exponent because of the
normalization \Ref{norm-scaled}.

\subsubsection*{Model IMM-A}
To test whether the Fourier transformed Boltzmann equation
\Ref{ft-t-eq} admits a scaling solution with a dominant small-$x$
singularity $x^a$, we make for the small$-x$ ansatz,
\be \label{ansatz}
\Phi(x) =  1-x -A x^a  ,
\ee
insert this expression in \Ref{ft-eq}, and investigate whether the
resulting equation admits a solution for the exponent $a$. This
is done by equating the coefficients of equal powers of $x^s$ on
both sides of the equation, which yields for general
dimensionality,
\ba \label{lambda-a}
\gamma &=& \lambda_1 \equiv \int_\bn[1-e_+(\bn) - e_-(\bn)]=
2p(1-p)/d
\nn a\gamma &=& \lambda_a \equiv \int_\bn [1-e_+^{a}(\bn)-e_-^{a}(\bn)] .
\ea
The eigenvalue $\lambda_s (s=1,a)$ has been calculated in
\Ref{A3} and \Ref{B7} of the Appendix. The relevant properties
are: {\em (i)} $\lambda_0 =0$ because of particle conservation;
{\em (ii)} $\lim_{s \to 0} \lambda_s =-1$; {\em (iii)} $\lambda_s$
is a concave function, monotonically increasing with $s$, and {\em
(iv)} all eigenvalues for non-negative {\em integers} $n$ are
positive (see Fig.1). In one dimension the above eigenvalue
becomes $\lambda_a=1-p^{2a}-(1-p)^{2a}$.

\begin{figure}[h]\label{fig1}
$$\psfig{file=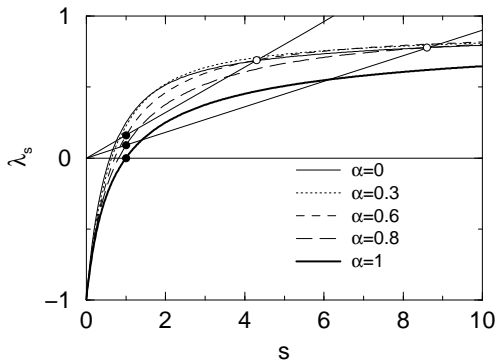,height=5cm,angle=270}$$
\caption{Eigenvalue $\lambda_s$ is a concave function of $s$,
plotted for different values of $\alpha$ for the 2-D inelastic
Maxwell model IMM-A. The line $y=s\lambda_1$ is plotted for
$\alpha=0.6,0.8$ and $\al =1$(top to bottom). The intersections
with $\lambda_s$ determine the points $s_0$ (filled circles) and
$s_1$ (open circles). Here $s_1=a$ determines the exponent of the
power law tail. For the elastic case ($\al=1, \lambda_1=0$, energy
conservation) there is only 1 intersection point. The branch of
$\lambda_s$ in the interval $s_0 < s < s_1$ is referred to as {\em
stable} and the branches $s<s_0$ and $s>s_1$ as {\em unstable}.}
\end{figure}

To continue we combine both relations in \Ref{lambda-a}, which
determine the exponent $a$ as the root of the {\em
transcendental} equation,
\be \label{trans-eq}
\lambda_s=s \lambda_1 .
\ee
This equation  has been solved numerically, and the results are
plotted in Fig.2. We note here that Krapivsky and Ben-Naim
\cite{BN+PK-6-11} have derived the same transcendental equation.

As can be seen from the graphical solution in Fig.1, the
transcendental equation \Ref{trans-eq} has two solutions, the
trivial one $(s_0=1)$ and the solution $s_1=a$ with $a>1$. The
numerical solutions for $d=2,3$ are shown in Fig.2a as a function
of $\alpha$, and the $\alpha$-dependence of the root $a(\alpha)$
can be understood from  the graphical solution in Fig.1.
 In the elastic limit as $\alpha\uparrow 1$   the eigenvalue
$\lambda_1(\alpha)\to 0$ because of energy conservation. In that
limit the transcendental equation \Ref{trans-eq}, $\lambda_s(1)-
s \lambda_1(1)=0$, no longer has a solution with $a>1$, and
$a(\alpha)\to\infty$, as it should be. This is consistent with a
Maxwellian tail distribution in the elastic case. Needless to say
that the transcendental equation for the one-dimensional IMM-A
has the solutions $s_0=1$ (trivial) and $s_1=a=3/2$, describing a
power law tail $\tilde{f}(c) \sim 1/|c|^{2a+d} =|c|^4$, in full
agreement with the  exact scaling solution $\tilde{f}(c) =
(2/\pi) (1+c^2)^{-2}$, found by Baldassarri et al.
\cite{Baldassarri-1D} for this case. The one-dimensional case is
a bit pathological because the intersection points
$(s_0,s_1)=(1,3/2)$ of $y=\lambda_s(\alpha)$ and
$y=s\lambda_1(\alpha)$ are independent of $p=\half (1+\al)$,
implying that these points are common to all $\lambda -$curves at
different parameter values $\al$. In summary, we conclude that
there exists for inelastic Maxwell models a scaling solution
$\tf(c)$ with a power law tail $ 1/|c|^{2a+d} $  at large
energies.

As a parenthesis we apply the previous analysis for similarity
solutions to the {\em elastic} case $(\alpha =1)$, where the BKW
- mode, discussed in the introduction, is an exact similarity
solution. In the elastic case similarity solutions of
\Ref{ft-t-eq} would also have the form $\phi(x,t) =
\Phi(\varepsilon_o(t)x)$, but the time dependent factor
$\varepsilon_o(t)$ is not determined by the {\em conserved}
second moment $\av{v^2} =\half d$.  So  the typical time scale
$\gamma$, entering through $\varepsilon_o(t)= \exp(-\gamma t)$,
will be determined by the lowest moment with $t-$dependence, i.e.
$\av{v^4}\sim \varepsilon_o^2(t)$, and its rate equation imposes
$\lambda_2=2\gamma$. The transcendental equation becomes then
$\gamma_s =\lambda_s - s\lambda_2/2$, which has again two
solutions, $s_0=2$ and $s_1=3$ because $\lambda_2/2=\lambda_3/3$.
As both $a$-values are {\em integers}, the small$-x$ behavior of
the characteristic function contains only regular terms $x^2$ and
$x^3$, which do not result in any power law tails. In fact, the
solutions $\{s=2,3\}$ correspond to the exact closed form
solution $\phi(x)=e^{-x}(1+x)$, the well-known Bobylev-Krook-Wu
mode for elastic Maxwell molecules \cite{bobylev-bgk,phys-rep-ME}.

\begin{figure}[h]\label{fig2}
$$\psfig{file=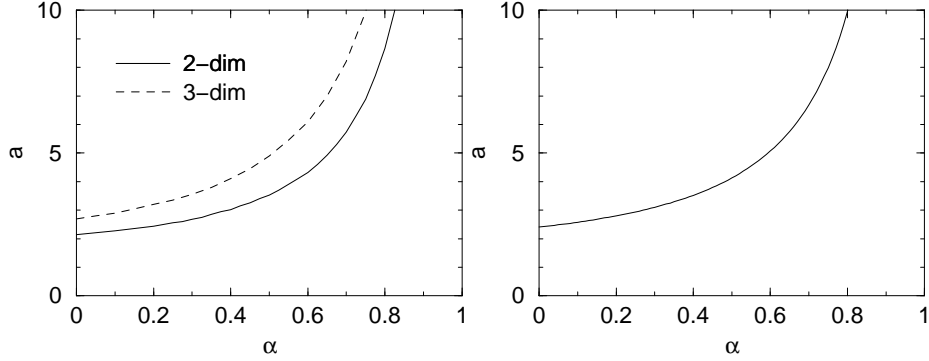,height=5cm,angle=270}$$
\caption{Exponent $ a(\al)$, which is the root of \Ref{trans-eq},
as a function of the coefficient of restitution $\alpha$, which
determines the high energy tail $1/c^{2a+d}$ of the scaling
solution $\tf(c)$. Left panel for IMM-A with uniform angle of
incidence in 2-and 3-D, and right panel for IMM-B with uniform
impact parameter in 3-D.}
\end{figure}

\subsubsection*{Model IMM-B}
The second part of this section deals with the Boltzmann equation
\Ref{IMMB} for model IMM-B with uniform impact parameters, as
introduced by Bobylev et al. \cite{bobylev}, and we show that the
above method gives similarity solutions with power law tails for
this model as well. We restrict the analysis to the
three-dimensional case. The integral equation for the
characteristic function in this case has been derived in
\cite{bobylev}, and reads,
\be \label{ft-eq-B}
\partial_t \phi(\bk,t)=  \int^\prime_\bn \;|\widehat{\bk} \cdot \bn|
\phi(\bk_+,t)\phi(\bk_{-},t)-\phi(\bk,t).
\ee
In a similar way as in the previous section, we derive the
integral equation for the corresponding scaling solution,
$\phi(x,t) =\Phi(x e_o(t))$ with the result,
\be \label{tf-eq-B}
-\gamma x \Phi'(x) +\Phi(x)=  \int^\prime_\bn
|\widehat{{\bk}}\cdot \bn| \Phi(xe_+)\Phi(xe_-),
\ee
and the procedure described in \Ref{ansatz} and \Ref{lambda-a}
leads to the same transcendental equation \Ref{trans-eq} with
eigenvalue,
\ba \label{lambda-s-B}
\lambda_a &=& \int^\prime_\bn \; |\widehat{{\bk}}\cdot \bn|
[1-e_+^{a} - e_-^{a}]
\nn &=& 2 \int^1_0 dx \,x \{ 1-p^{2a} x^{2a} -[1-z x^2]^{a}\}
\nn &=& 1 - \frac{ p^{2a}}{a+1} -\frac{1}{a+1}
\left\{\frac{1-(1-p)^{2a+2}}{1-(1-p)^{2}} \right\}.
\ea
The equality on the second line has been obtained by changing to
the new integration variable $x= \cos \theta$, and the value for
$\lambda_{a}$ is in agreement with the corresponding eigenvalue
obtained in \cite{bobylev} for integer $a$. For $a=1$ one obtains
here $\lambda_1 = p(1-p)$, in agreement with \Ref{2-moment} for
d=3. The numerical solution of the transcendental equation
\Ref{trans-eq} is shown in Fig.2b.

It turns out that the numerical values $a(\al)$ for the exponents
of IMM-B with a uniform distribution of impact parameters looks
qualitatively the same as those for model IMM-A with a uniform
distribution of angles of incidence.   The differences in
distribution of impact parameters of both models, uniform in IMM-B
versus biased towards grazing in IMM-A, has no qualitative effect
on the nature of the singularity, i.e. on the values  or
$\al$-dependence of the tail exponents. Therefore, the power law
tail in $\tilde f(c)\sim1/c^{2a+d}$ is  not a spurious effect
induced by models with impact parameters  $b$ biased towards
grazing collisions.

After completion of this article Bobylev and
Cercignani\cite{B+C-priv-com} have kindly informed us that
equations \Ref{ft-eq-B} and \Ref{tf-eq-B} are also correct for
the $d-$dimensional version of model IMM-B. This implies not only
that the exponent $a$ can be calculated for general
dimensionality, but it also implies that the next section, after
some trivial modifications, applies to the model IMM-B in $d$
dimensions as well.

We conclude this section by discussing some {\em numerical}
evidence for the conjecture on the approach to a scaling solution
with algebraic high energy tails, as formulated in section I.
Baldassarri \cite{baldassarri-private} has obtained large-$t$
solutions $f(v,t)$ by applying the DSMC (Direct Simulation Monte
Carlo) method to the Boltzmann equations for three types of
inelastic Maxwell models, among which the two-dimensional IMM-A
model and the three-dimensional IMM-B model, analyzed in this
article. For the totally inelastic case ($\alpha=0$) he has
observed that $f(v,t)$, for general initial data, evolves after
sufficiently long time to a scaling solution of the form
\Ref{def-f-ct}, on which the simulation data can be collapsed.
Moreover it has a power law tail,  $\tf (c) \sim 1/c^{2a+d}$ in
agreement with the predictions in Figs. 2a,b at $\al =0$.

\renewcommand{\theequation}{IV.\arabic{equation}}
\setcounter{section}{3} \setcounter{equation}{0}
\section{Approach to Scaling Solutions}

\subsection{Moment Equations}

The moment equations for Maxwell models are special because they
form a closed set of equations that can be solved sequentially as
an initial value problem. In this section we study the effects of
power law tails $\tilde f(c)\sim A/c^{2a+d}$ on the moments, and
investigate in what sense, if any, the calculated time dependence
of the moments as $t\to\infty$ is related to the singular behavior
\Ref{ansatz} of the scaling form $\Phi(x)$, derived in section
III. The latter implies that the moments $\mu_n$, generated by
$\Phi(x)$, with $n>a$ are {\em divergent} and that those with
$n<a$ remain {\em finite}.

First consider the standard moments $m_n(t)$. Inserting the
expansion \Ref{mom-t-exp} in \Ref{ft-t-eq} and equating the
coefficients of equal powers of $x$ yields for the moments
$m_n(t)= \langle v^{2n}\rangle_t /(d/2)_n$ the following equations
of motion,
\be \label{A2}
\dot{m}_n + \lambda_n m_n = {\sum^{n-1}_{l=1}}H(l,n-l) m_l
m_{n-l},
\ee
where the coefficients $H(l,m)$ and eigenvalues $\lambda_n$ for
model IMM-A are defined and calculated in \Ref{A3}-\Ref{B7}.
Those for model IMM-B follow after some trivial replacement,
indicated in Appendix A. Regarding the moments, we have $m_0=1$
because of \Ref{norm}, and $m_1(t)=\exp (-\lambda_1t)m_1(0)$,
where $\lambda_1 =\gamma$, as given in \Ref{2-moment}. Moreover as
$t\to \infty$, all moments with $n>0$ vanish, which is consistent
with the limiting behavior $f(v,t)\to \delta^{(d)}(\bv)$ for
$t\to\infty$.

Next we consider the moments $\mu_n$ generated by the scaling form
$\Phi(x)$. Using a self-consistent argument we have demonstrated
in  section III that the kinetic equation \Ref{IMMA} admits a
scaling solution $\Phi(x)$ with a dominant small-$x$ singularity
$x^a$, with $a>1$ and non-integer. This implies that all $n-$th
order derivatives of $\Phi(x)$ at $x=0$, or equivalently all
moments $\mu_n$, are {\em finite} if $n \leq n_o =[a]<a$, and all
those with $n>a$ are {\em divergent}. Here $[a]$ is the largest
integer less than $a$. Hence, the {\em small}-$x$ behavior of
$\Phi(x)$ can be represented  as ,
\be \label{A4}
\Phi(x) =   {\sum_{n=0}^{n_o}} (-x)^n \frac{\mu_n}{n!}
 + {o}(x^{a}),
\ee
where the remainder is of order ${o}(x^{a})$  as $x \to 0$. In
this scaling form we only know the exponent $a$ and the moments
$\mu_0=\mu_1=1$. Now we calculate the unknown finite moments of
the scaling form, $\mu_n$ with $1< n <a$. This is done by
inserting \Ref{A4} into the kinetic equation \Ref{ft-t-eq},
yielding the recursion relation,
\ba \label{A5} \mu_n &=&
(1/\gamma_n) \sum_{l=1}^{n-1} H(l,n-l) \mu_l \mu_{n-l} \nn
\gamma_n &=& \lambda_n-n\gamma =\lambda _n -n\lambda_1.
\ea
Here $\gamma_1\equiv \gamma-\lambda_1=0$ on account of
\Ref{2-moment} and the initialization is $\mu_1=1$. The solutions
$\mu_n$ for $n=2,3,4,5$ in model IMM-A are shown in Fig.3 as a
function of the coefficient of restitution $\alpha$. Furthermore
we observe that the root $s=a$ of the transcendental equation
\Ref{trans-eq}, $\gamma_s =\lambda_s-s\lambda_1=0$, indicates
that   $\gamma_s$ changes sign at $s_1=a$ (see open circles in
Fig.1), and that according to Section III all moments $\mu_n$
with $n>s_1=a$ ($n$ on unstable branch) are {\em divergent}.

\begin{figure}[h]\label{fig3}
$$\psfig{file=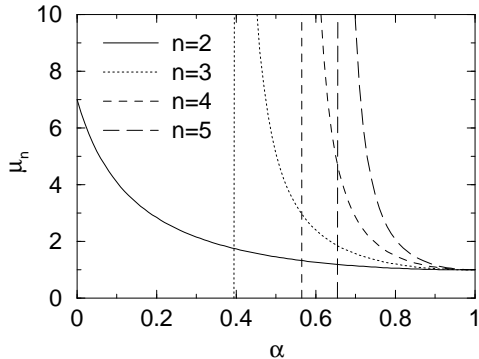,height=5cm,angle=270}$$
\caption{The moments $\mu_n \sim \av{c^{2n}}$ of the scaling form
$\tf(c)$ for $n=2,3,4,5$ as a function of $\alpha$. The moment
$\mu_2$ exists for all values of $\alpha$, while higher moments
only exist for $\alpha$ above a threshold value, indicated by the
vertical asymptotes.}
\end{figure}

The recursion relation \Ref{A5} for the moments $\mu_n$ in the
one-dimensional case is again a bit pathological in the sense
that the {\em stable} branch $(s_0=1 < s < s_1=3/2)$ contains
only one single integer label, i.e. $s=1$. So only $\mu_0=\mu_1=1$
are finite, and all other moments are infinite, in agreement with
the exact solution of Baldassarri et al.

The recursion relation \Ref{A5} for both models has a {\em
second} set of solutions $\{ \mu_n \}$, which has been studied by
Bobylev et al. \cite{bobylev}, who showed that within this set
there are  moments $\mu_n$ being {\em negative}. The argument is
simple. Consider $\mu_n$ in \Ref{A5} with $n=n_o+1$. Then the
prefactor $1/\gamma_{n_o+1}$ on the right hand side of this
equation is negative because the label $n_o +1 >a$ is on the
unstable branch of the eigenvalue spectrum in Fig.1, while all
other factors are positive. This implies that the corresponding
scaling form $\tf (c)$  has negative parts, and is therefore
physically not acceptable. We also note that the moments $\mu_n$
of the physical and the unphysical scaling solution $\Phi(x)$
coincide as long as both are finite and positive in the
$\al-$interval  that includes $\alpha =1$.

\subsection{Long time behavior of rescaled moments}

In the introduction we have already indicated that the
distribution function $f(v,t)$ itself approaches a Dirac delta
function, $\delta^{(d)} (\bf v)$, and that the rescaled
distribution function $\tf(c,t) = v_o^d(t) f(v_o(t)c,t)$, as
defined in \Ref{def-f-ct}, supposedly approaches a scaling form.
To demonstrate how this happens, we define the rescaled function
$\tilde\phi(x,t)$ through the relation $\phi(x,t)= \tilde
\phi(e_o(t)x,t)$ with $e_o(t)=\exp[-\gamma t] e_o(0)$. We also
consider the more general case where the typical time scale is not
a priori fixed by defining $e_o(t)$ in terms of one specific
moment $\av{v^{2n}}$. Then it follows from \Ref{ft-t-eq} that
\be\label{A6}
\partial_t \tilde\phi(x,t)-\gamma x\partial_x\tilde\phi(x,t)
+ \tilde\phi(x,t) = \int_\bn \tilde\phi(xe_+,t)\tilde\phi(xe_-,t),
\ee
where $\tilde{\phi}(x,t)$ with $x=\fourth k^2$ is the Fourier
transform of $\tf(c,t)$ in \Ref{def-f-ct}. The stationary
solution of this equation determines a possible similarity
solution for a given $\gamma$. The equations of motion for the
rescaled moments $\mu_n(t)$ of $\tilde f(c,t)$ are obtained by
substituting the Taylor expansion $\tilde\phi(x,t) = \sum_n
(-x)^n \mu_n(t)/n!$, into \Ref{A6}, which yields for $n=1,2,\dots$
\ba \label{A7}
\dot\mu_n +\gamma_n \mu_n &=& \sum_{l=1}^{n-1} H(l,n-l)
\mu_l\mu_{n-l} \nn \gamma_n  &=& \lambda_n -n\gamma ,
\ea
with  $\mu_n(0)=m_n(0)/e^n_o(0)$. We also observe that the
equations for $\tilde{\phi}(x,t)$ and for the moments $\mu_n(t)$
hold for arbitrary rescaled initial distributions, $ \tf(c,0) =
v^d_o(0) f(c v_o(0),0)$.

\begin{figure}[h]\label{fig4}
$$\psfig{file=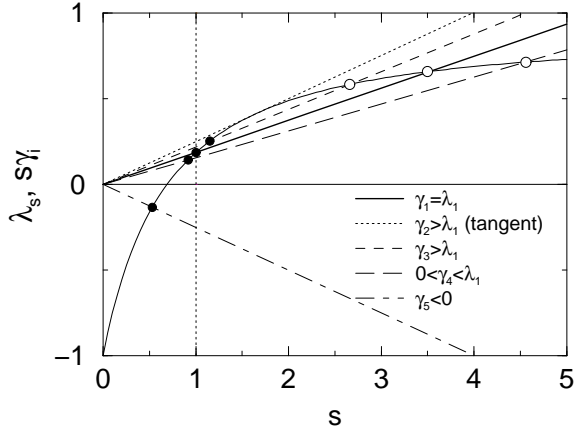,height=6cm,angle=270}$$
\caption{Graphical construction of the solutions of $\lambda_s=
s\gamma$ for different values of $\gamma$ at dissipation
$\alpha=0.5$.  Intersections $(s_0)$ with smaller $s-$values are
marked with filled circles, and intersections $(s_1)$ with larger
$s-$values  with open ones. }
\end{figure}

To study the long time approach of $\tilde f(c,t)$ to the scaling
solution $\tf(c)$ in \Ref{def-f-ct}, we analyze the approach of
the rescaled  moments $\mu_n(t) = m_n(t)/e^n_o(t)$ to the moments
$\mu_n$ of the scaling solution $\tf(c)$. To do so we have to
choose $\gamma=\lambda_1$ (see Fig.4) on account of \Ref{norm} and
\Ref{2-moment}. The infinite set of moment equations \Ref{A7} for
$\mu_n(t)$ can be solved sequentially for all $n$ as an initial
value problem. To explain what is happening, it is instructive to
use a graphical method to determine the zeros of the eigenvalue
$\gamma_s=\lambda_s-s\gamma$ for different values of $\gamma$.
This is illustrated in Fig.4 by determining the intersections
$\{s_0,s_1\}$  of the curve $y=\lambda_s$ and the line
$y=s\gamma$, where $s_0$ and $s_1$ are denoted respectively by
filled ($\bullet$) and open circles ($\circ$). These circles
divide the spectrum into a (linearly) {\em stable} branch ($s_0
<s<s_1$) and two {\em unstable} branches ($s<s_0$) and ($s>s_1$).
The moments $\mu_s(t)$ with $s=n
>a$ are on an unstable branch $(\gamma_s <0)$ and will grow for
large $t$ at an exponential rate, $ \mu_n(t) \simeq \mu_n(0)
\exp[|\gamma_n|t]$, as can be shown by complete induction from
\Ref{A7} starting at $n=[a] +1$. They remain positive and finite
for finite time $t$, but approach $+ \infty$ as $t\to\infty$, in
agreement with the predictions of the self consistent method of
Sect.III. The moments  with $n$ on the stable branch are linearly
stable ($\gamma_n>0$), but may still grow through nonlinear
couplings with lower moments $\mu_l(t)$ whenever $l=1,2,\cdots$
is on the unstable branch ($l<s_0$). This is relevant for the
discussion in the next subsection. In the case $\gamma
=\lambda_1$ under consideration however, all moments $\mu_l$ with
$l=1,2, \cdots, [a]$ are globally stable and approach for
$t\to\infty$ the limiting value $\mu_n(\infty)=\mu_n$,  which are
the finite positive moments of the scaling form \Ref{A4}, plotted
in Fig.3.

The behavior of the moments described above is considered as a
weak form of convergence or approach of $\tf(c,t)$ to $\tf(c)$
for $t \to \infty$. This result will be described in a more
precise form and summarized in the last section. The physically
most relevant distribution functions $f(v,t)$ and rescaled
distribution functions $\tf(c,t)$ are those with {\em regular}
initial conditions, i.e. all moments $m_n(0)= e^n_o(0) \mu_n(0) <
\infty$. This implies that the initial condition for the rescaled
Fourier transform $\tphi(x,0)$ with $x=\fourth k^2$ of $\tf(c,0)$
has a series expansion regular at the origin $x=0$, i.e. all its
derivatives exist in that point.

\subsection{ More similarity solutions}

The results of the previous subsection can also be generalized to
different values of $\gamma$ and different classes of initial
distributions, which possess only a finite set of bounded moments
$m_n(0)$. Here we summarize only the most important results of
these generalizations, the details of which will be published
elsewhere \cite{ME+RB-levy}.

The case, $\gamma>\lambda_1$, concerns the line $y= s \gamma_2$ in
Fig.4. Then $\mu_1(t)$ will always be in the unstable region
(there is either an intersection point $s_0>1$, or there are no
intersection points at all). So, $\mu_1(t)\to\infty$ as
$t\to\infty$ at an exponential rate, and it drives all moments
$\mu_n(t) \to\infty$ through nonlinear couplings to $\mu_1(t)$,
which is present on the right hand side of \Ref{A7}.
Consequently, there is no approach to the scaling form $\Phi(x)$
with a rate constant $\gamma>\lambda_1$.

The case, $0<\gamma<\lambda_1$, concerns the line $y=s \gamma_4$
where there are two intersection points $\{s_0,s_1\}$ with
$s_0=a<1$ and $s_1=b>1$, corresponding to the dominant
singularity $x^a$ and a subleading singularity $x^b$. The
dominant singularity corresponds to a power law tail $\tilde
f(c)\sim 1/c^{2a+d}$ with $a<1$. Because $a<1$, the system has
infinite energy at {\em all} times. All higher moments $\mu_n$
with $n=2,3,\dots$ are {\em divergent} as well at all times.
Consequently, the initial states under discussion are already
singular with $\phi(x,0)\simeq 1-A x^a$. Of course such states
are of much less interest for possible physical applications than
the regular ones, discussed in the previous subsection. The
feature of interest here is to demonstrate that inelastic
Boltzmann equations generate for initial states characterized by a
singularity $x^a$ with $a<1$ a new type of singularity $x^b$,
which is found through the graphical construction using line $y=s
\gamma_4$. For elastic Maxwell molecules such states have been
analyzed recently  by Bobylev and Cercignani
\cite{AB+CCpreprint}. In one-dimensional systems these initial
states are closely related to Levy distribution \cite{Montroll}
with the characteristic function $\varphi(k)=\exp[-bk^{2a}]$,
where $b$ is positive. For such distributions it is well known
that Fourier inversion yields for $0<a<1$ a {\em non-negative}
distribution $f(v)$ with a power law tail $1/v^{2a+d}$ with
$d=1$. On the other hand, for $a>1$ Fourier inversion may lead to
a distribution with negative $f(v)$.

However, in this case one can say more. Following Bobylev and
Cercignani, we assume that we can construct a {\em non negative}
initial distribution with a $\phi(x,0)$ which is a regular
function of $\zeta=x^a$ in a neighborhood of $\zeta=0$, i.e.
\be \label{star}
\phi(x,0) = \sum_{n=0}^ \infty (-1)^n \frac{x^{na}}{n!} A_n(0),
\ee
where $|A_n(0)|<A^n$, and we have chosen normalization such that
$A_0(0)=A_1(0)=1$. We have slightly modified the example of
\cite{AB+CCpreprint} to have a finite non-vanishing radius of
convergence of \Ref{star}. Then we can demonstrate that the
rescaled characteristic function $\tilde\phi(y,t) \equiv
\phi(x,t)$ approaches in the scaling limit ($t\to\infty$ with
$y\equiv x\exp[-\gamma t]$ fixed) a positive scaling form or
similarity solution $\Phi(y)$ with $\gamma = \lambda_a/a$, and
\be\label{starstar}
\Phi(y)=\sum_{n=0}^{[b]} (-1)^ n \frac{x^{na}}{n!} A_n(\infty)
+o(y^b),
\ee
where $A_0(\infty)=A_1(\infty)=1$. The approach is again in the
weak sense of a finite set of moments. The $A_n(\infty)$ are
positive and can be calculated from a set of recursion relations,
rather similar to \Ref{A5}. Moreover, $b=s_1$ is the left most
intersection point of $\lambda_s$ with the line $y=s\gamma_5$
(see the left most open circle on $\lambda_s$). So, the initial
$\phi(x,0)$, which is regular in $\zeta=x^a$ around $\zeta=0$,
develops a {\em new} singularity of the type $x^b$. This can
again be demonstrated by considering the rescaled function $\tilde
\phi(y,t)$, defined as $\phi(x,t)=\tilde \phi(e^{-\gamma t}y,t)$,
and expanding $\tilde\phi(y,t)$ in a series like \Ref{star} with
$A_n(0)$ replaced by $A_n(t)$. The coefficients satisfy moment
equations, rather similar to \Ref{A7}.

In the case, $\gamma\leq 0$, the results are similar to those in
the previous paragraph, except that there is only one intersection
point at $s_0=a$ (see the line $y=s\gamma_5= -|\gamma_5 |s$). The
energy and all higher moments are infinite, and the scaling form
of the characteristic function  $\Phi(x)$ is a {\em regular}
function of $\zeta = x^a$ near the origin. A similar solution for
the elastic case has been obtained in Ref. \cite{AB+CCpreprint}.

\section{Conclusions}

Using self consistency arguments we have shown that the nonlinear
Boltzmann equation for the inelastic Maxwell models IMM-A and
IMM-B admits a scaling solution $\tf(c)$ with a power law tail
$1/c^{2a+d}$ where the exponent $a$ is given by the root $a$ with
$a>1$ of a transcendental equation. This implies that all moments
$\mu_n$ of $\tf(c)$ are divergent for $n>a$, and those $\mu_n$
with $n<a$ are positive and finite, and are given recursively
through \Ref{A5} with initialization $\mu_1=1$.

 For systems with dissipative dynamics we have formulated a
conjecture about the long time approach, for general classes of
initial distributions, of the rescaled distribution function
$\tf(c,t) = v^d_o(t) f(c v_o(t),t)$  to a scaling solution
$\tf(c)$ with an  high energy tail, which is overpopulated when
compared to a Gaussian distribution. For the inelastic Maxwell
models, studied in detail in this article, we have demonstrated
this approach to the scaling solution $\tf(c)$ in subsection IV.B
in the following weak sense:\\
Given on the one hand the rescaled moments $\mu_n(t)
=m_n(t)/e^n_o(t)$ of the physically most relevant class of {\em
regular} initial distributions with all moments $m_n(0)$ bounded,
and rescaled according to \Ref{def-f-ct}, and given on the other
hand the unique set of finite positive moments $\mu_n $  of the
scaling form with $n<a$,  then all $\mu_n(t)$ with $n<a$ approach
for $t \to \infty$ the limiting value $\mu_n$ through a sequences
of positive numbers, and all moments $\mu_n(t)$ with $n>a$ behave
as $\mu_n(t) \simeq \mu_n(0) \exp[|\gamma_n|t] \to +\infty$ as $t
\to \infty$. This is in agreement with all known properties of
the scaling solution.

 We consider the long time behavior of the
set of moments $\mu_n(t) \to \mu_n \;\;(n<a)$ and $\mu_n(t) \to
+\infty \;\;(n>a)$ as a demonstration that $\tf(c,t)$ approaches
 the scaling form $\tf(c)$ for $t \to \infty$. We
further note that these results for the long time behavior of
inelastic Maxwell models, after the rescaling of the initial
distribution, are {\em universal}, i.e. they are independent of
all details of the initial distributions.  We consider the present
results for inelastic Maxwell models as strong support for our
more general conjecture, which is also confirmed for inelastic
Maxwell models  through the Monte Carlo  simulations of the
Boltzmann equation by Baldassarri et al, in which the approach of
$\tf(c,t)$ to a positive power law tail with the predicted
exponent $a$ was confirmed within  reasonably small error bars.

\setcounter{equation}{0}
\renewcommand{\theequation}{A.\arabic{equation}}
\section*{Appendix A: Angular averages in $\mbox{d}$-dimensions}

The angular average \Ref{mom-t-n}  of powers of $\hat{\bf
a}\cdot\bn =\cos \theta$ can be simply calculated by using polar
coordinates with $\hat{\bf a}$ as the polar axis. Then,
\be \label{B1}
\beta_{s} =\int_\bn |\hat{\bf a}\cdot\bn|^{s} =
\frac{\int_0^{\pi} d\theta (\sin \theta)^{d-2} (\cos\theta)^{s} }
{\int_0^{\pi} d\theta (\sin\theta)^{d-2}}
=\frac{\Gamma(\frac{s+1}{2}) \Gamma(\frac{d}{2})}
       {\Gamma(\frac{s+d}{2}\Gamma(\frac{1}{2}) }.
\ee
For $s=2n$ this formula can be expressed as
\be \label{B2}
\beta_{2n} = (1/2)_n/(d/2)_n,
\ee
where the Pochhammer symbol is defined as
\be\label{B3}
(a)_n= \Gamma(a+n)/\Gamma(a)= a(a+1)\dots a (a+n-1).
\ee
In fact, we will use the notations $(a)_s$, $\beta_{2s}$, $
s!=(1)_s$ and $\xovery{s}{r}$ also for non-integer values of $s$
by expressing these quantities in terms of Gamma functions.

Next we define $H(l,m)$ and $\lambda_n$, introduced in \Ref{A2}
for model IMM-A,
\ba
\label{A3}
& H(l,n-l)= \xovery{n}{l}\displaystyle \int_\bn
e_+^l(\bn) e_-^{n-l}(\bn) &
\nn &\lambda_n = 1 + \delta_{n0}-H(0,n)-H(n,0)= \displaystyle \int_\bn
[ 1 +\delta_{n0} - e_+^n(\bn) - e_-^{n}(\bn) ],&
\ea
where $e_\pm(\bn)$ has been defined in \Ref{k-var}. For model
IMM-B one needs to replace $\int_\bn$ with $\int^\prime_\bn
|\hat{a} \cdot \bn|$. These expressions hold for $d=1,2,....$.
Evaluation of $H(l,m)$ requires,
\be\label{B4}
I(m,n)=\int_\bn e_+^m e_-^n =\frac{2
p^{2m}}{B(\frac{d-1}{2},\frac{1}{2})} \int _0^{\pi/2} d\theta
(\sin\theta)^{d-2} (\cos\theta)^{2m} [1-z\cos^2\theta]^n,
\ee
where $z=1-(1-p)^2$. Following Krapivsky and Ben-Naim
\cite{BN+PK-6-11} we change to the new integration variable
$\mu=\cos^2\theta$, to find for $d=1,2,\dots$
\ba \label{B5}
H(m,n) &=& \xovery{n+m}{m}
\frac{p^{2m}}{B(\frac{d-1}{2},\frac{1}{2})} \int_0^1 d\mu
\mu^{m-1/2} (1-\mu)^{\frac{d-3}{2}} [1-z\mu]^n \nn &=&
\xovery{n+m}{m} \beta_{2m} p^{2m} \,_2F_1
(-n,m+\frac{1}{2},m+\frac{d}{2};z) \nn &=& p^{2m} \xovery{n+m}{m}
\sum_{l=0}^n\xovery{n}{l} \beta_{2l+2m} (-z)^l.
\ea
On the second and third line we have used the fundamental
integral representation for the hyper-geometric function $\
_2F_1$, and its Gauss series, i.e.
\ba \label{B6}
\,_2F_1(a,b,c;z)= \left\{
\begin{array}{l} B^{-1}(b,c-b) \int_0^1 dt
t^{b-1}(1-t)^{c-b-1} (1-zt)^{-a} \\[2mm]
\sum_{l=0}^\infty \frac{(a)_l(b)_l}{(c)_ll!} z^l.
\end{array} \right.
\ea
When $a=-n,\, (n=0,1,2\dots)$, then $\,_2F_1(-n,b,c;z)$ is a
polynomial of degree $n$ in $z$, and the Gauss series ends at
$l=n$.

To calculate the eigenvalue $\lambda_n$ in \Ref{A3} we deduce
from \Ref{B6}
\ba\label{B7}
 H(n,0) &=& \beta_{2n} p^{2n} \nn
 H(0,n) &=&
\,_2F_1 (-n,\frac{1}{2},\frac{d}{2};z)= \sum_{l=0}^n
\xovery{n}{l} \beta_{2l} (-z)^l .
\ea

\section*{Acknowledgements}
The authors want to thank A. Baldassarri et al for making their
simulation results available to them,  and C. Cercignani, A. Bobylev
and A. Santos for helpful correspondence. M.E. wants to thank E.
Ben-Naim for having stimulated his interest in dissipative
one-dimensional Maxwell models during his stay at CNLS, Los
Alamos National Laboratories in August 2000.
This work is supported by DGES (Spain), Grant No BFM-2001-0291.
Moreover R.B.
acknowledges support of the foundation "Fundamenteel Onderzoek
der Materie (FOM)", which is financially supported by the Dutch
National Science Foundation (NWO).

\end{document}